\documentclass[prd,preprint,superscriptaddress,preprintnumbers,eqsecnum,showpacs,nofootinbib,nobibnotes]{revtex4}
\usepackage{amsfonts,amsmath,amssymb,bm,natbib}
\usepackage{graphicx} 

\newcommand{\be}{\begin{equation}}
\newcommand{\bea}{\begin{eqnarray}}
\newcommand{\ee}{\end{equation}}
\newcommand{\eea}{\end{eqnarray}}

\def\spr{\!\cdot\!}

\def\s#1{{\scriptscriptstyle #1}}

\def\noeq#1{(\ref{#1})}
\def\1eq#1{Eq.~(\ref{#1})}

\def\2eqs#1#2{Eqs.~(\ref{#1}) and~(\ref{#2})}
\def\3eqs#1#2#3{Eqs.~(\ref{#1}),~(\ref{#2}) and~(\ref{#3})}

\def\fig#1{Fig.~\ref{#1}}

\def\diff{{\rm d}}
\def\gA{g^2 C_A}

\def\ie{{\it i.e.}, }
\def\eg{{\it e.g.}, }

\def\cd{\!\cdot\!}

\def\dE#1{{#1}_\s{\rm{E}}}

\def\Dxi{D}

\def\Deltaxi{\Delta}

\def\DeltaL{\Delta_\s{\mathrm{L}}}

\def\FL{F_\s{\mathrm{L}}}
\def\GammaL{\Gamma^\s{\mathrm{L}}}

\def\m2L{m^2_\s{\mathrm{L}}}

\def\c{c}
\def\cSDE{c_\s{\mathrm{SDE}}}
\def\cNI{c_\s{\mathrm{NI}}}
\def\a{a}

\begin{document}

\title{Yang-Mills two-point functions in linear covariant gauges}

\author{A.~C. Aguilar}
\affiliation{University of Campinas - UNICAMP, 
Institute of Physics ``Gleb Wataghin'',
13083-859 Campinas, SP, Brazil}

\author{D. Binosi}
\affiliation{European Centre for Theoretical Studies in Nuclear
Physics and Related Areas (ECT*) and Fondazione Bruno Kessler, \\Villa Tambosi, Strada delle
Tabarelle 286, 
I-38123 Villazzano (TN)  Italy}

\author{J. Papavassiliou}
\affiliation{\mbox{Department of Theoretical Physics and IFIC, 
University of Valencia and CSIC},
E-46100, Valencia, Spain}

\begin{abstract} 

In this work we use two different but complementary approaches in order to study the ghost propagator  of a pure SU(3) Yang-Mills theory quantized  in the linear covariant  gauges, focusing on its dependence on the gauge-fixing  parameter $\xi$ in the deep infrared. In particular, we first solve the Schwinger-Dyson equation that governs the dynamics of the ghost propagator, using a set of simplifying  approximations, and under the crucial assumption that the gluon propagators for $\xi>0$ are infrared finite, as is the case in the Landau gauge $(\xi=0)$. Then we appeal to the  Nielsen identities, and express the derivative of the ghost propagator with respect to $\xi$  in  terms of  certain auxiliary  Green's functions, which are subsequently computed under the same assumptions as before. Within both formalisms we find  that for $\xi>0$ the ghost dressing function approaches  zero in the deep infrared, in sharp contrast  to  what happens  in  the Landau  gauge, where it known to saturate at  a  finite (non-vanishing) value. The Nielsen identities are then extended to the case of the gluon propagator, and the $\xi$-dependence of the corresponding gluon masses is derived using as input the results obtained in the previous steps.  The result turns out to be logarithmically divergent in the deep infrared; the compatibility of this behavior  with the basic assumption of a finite gluon propagator is discussed, and a specific Ansatz is put forth, which readily reconciles both features. 

\end{abstract}


\maketitle

\section{Introduction}

The infrared (IR) behavior of Yang-Mills Green's functions in the Landau gauge has been the subject of numerous studies in the past few years, both in the continuum and on the lattice. Particularly important in this challenging quest has been the two-point sector of the theory, where it has been firmly established~\cite{Cucchieri:2007md,Cucchieri:2010xr,Bogolubsky:2007ud,Bogolubsky:2009dc,Oliveira:2009eh} that the gluon propagator saturates in the deep IR, a behavior directly associated with the dynamical generation of a momentum-dependent gluon mass~\cite{Cornwall:1981zr,Bernard:1982my,Donoghue:1983fy,Philipsen:2001ip,Aguilar:2004sw,Aguilar:2011ux,Binosi:2012sj,Aguilar:2014tka}, and that the ghost propagator remains massless, being accompanied by a dressing function that reaches a finite value at the origin~\cite{Boucaud:2008ky,Aguilar:2008xm}\footnote{For additional studies and alternative approaches, see~\eg~\cite{Dudal:2008sp,Fischer:2008uz,Pennington:2011xs,Campagnari:2010wc} and references therein.}. Interestingly enough, these characteristic features persist when implementing the transition from pure Yang-Mills to real world QCD; specifically, the inclusion of a small number of dynamical light quarks induces quantitative but not qualitative changes to the gluon and ghost propagators~\cite{Sternbeck:2007ug,Bowman:2007du,Ayala:2012pb,Aguilar:2012rz,Aguilar:2013hoa}.

Given that the Green's functions depend on both the gauge-fixing scheme employed and the choice of the gauge fixing parameter (gfp), it is important to explore their main dynamical features in different gauges, in order to filter out the truly gauge-independent properties of the theory. In particular, it would be interesting to establish the extent of validity and the possible modifications induced to the underlying mechanisms that endow the fundamental degrees of freedom, namely quarks and gluons, with their corresponding dynamical masses. Furthermore, even though physical observables are ostensibly gauge-independent, nonperturbative calculations are subject to truncations, which in turn may distort the delicate conspiracy of terms that produce the required gauge cancellations. It would be therefore a useful exercise to probe explicitly the gauge-(in)dependence of certain special combinations of Green's functions that are extensively used in a variety of phenomenological applications~\cite{Maris:1999nt,Aguilar:2009nf,Aguilar:2010gm,Qin:2011dd,Chang:2009zb,Cloet:2013jya,Binosi:2014aea}.

Among the  different classes of  gauges, the linear covariant (or  $R_\xi$) gauges~\cite{Fujikawa:1972fe} hold a prominent position. The corresponding gauge-fixing term that must be added to the standard Yang-Mills Lagrangian is given by $\frac{1}{2\xi} (\partial^\mu A^a_\mu)^2$, where $\xi$ represents the gfp; some characteristic values include the aforementioned Landau  gauge ($\xi=0$) and the Feynman gauge ($\xi=1$).  $R_\xi$ gauges have the advantage of manifest Lorentz covariance, and are particularly easy to use in diagrammatic calculations. In addition, by using the novel algorithm proposed in~\cite{Cucchieri:2009kk}, they can be implemented in numerical simulations of lattice regularized Yang-Mills theories even for $\xi\neq0$~\cite{Cucchieri:2011pp}. 

In the present work we initiate a study of the IR dynamics of the Yang-Mills two-point functions within this latter class of gauges, with the main objective to go beyond the standard Landau gauge paradigm. To that end, we will resort to two distinct but complementary approaches: on the one hand the Schwinger-Dyson equations (SDEs)~\cite{Roberts:1994dr} of the theory, and on the other the so-called Nielsen identities (NIs)~\cite{Nielsen:1975fs,Nielsen:1975ph}.

Within the SDE context, we focus exclusively on the integral equation governing the dynamics of the ghost dressing function, $F(q^2)$, which has a much simpler structure than the corresponding equation for the gluon propagator.At the formal level, the SDE in question is written down for general $\xi$, and after approximating the ghost-gluon vertex by its tree-level value, the solutions are obtained for the range $0<\xi \le 1$, thus spanning the values between the Landau and the Feynman gauges.Our main finding is that, contrary to what occurs in the Landau gauge, $F(q^2)$ vanishes as $q^2\to 0$ for all values of $\xi$  within the aforementioned interval. This drastic change in the infrared behaviour of $F(q^2)$ away from the Landau gauge may be traced back to the massless contributions associated with the $\xi$-dependent part of the gluon propagator entering into the ghost SDE. Specifically, even if one assumes that the cofactor $\Delta(q^2)$ of the transverse part of the gluon propagator is finite in the deep IR (as happens in the Landau gauge), it is a text-book fact that the longitudinal part (proportional to $\xi$) receives no quantum corrections, and maintains its  tree-level form [see \1eq{defprop}]. This massless contribution, in turn, introduces an infrared divergence into the ghost SDE, which, within the approximations employed, can be counteracted only if the solution for $F(q^2)$ vanishes in the deep IR. In particular, as we will see in detail, $F(q^2)$ vanishes at the very mild rate of $(-\c\,\xi\log q^2/\mu^2)^{-1/2}$ (with $\c>0$).

We then turn to the NIs, which express the gauge-dependence of ordinary Green's functions (propagators, vertices, etc.)  in terms of special auxiliary functions associated with the extended Becchi-Rouet-Stora-Tyutin (BRST) sector of the theory\footnote{The gfp-dependence of Green's functions can be in principle obtained also by using the so-called Landau-Khalatnikov-Fradkin (LKF) transformations~\cite{Landau:1955zz,Fradkin:1955jr}. 
These transformations have been used only in an Abelian context and are in general formulated in position space; therefore, their use for the problem at hand appears to be less direct.}. 
In the case of the ghost dressing function, the corresponding NI permits us to estimate its first derivative of $F(q^2)$ with respect to $\xi$, for arbitrary values of $\xi$; however, for practical purposes we limit our analysis to those $\xi$ that satisfy the condition $\xi\ll 1$. The reason for this choice is that, in this particular limit, the auxiliary functions appearing in the NI may be computed in their one-loop dressed approximation, using as input the gluon and ghost propagators known from the Landau gauge. The emerging expressions, when evaluated in the deep infrared, reproduce rather faithfully the behavior obtained from the ghost SDE; specifically, up to a multiplicative factor, one recovers precisely the derivative of $(-\c\,\xi\log q^2/\mu^2)^{-1/2}$ with respect to $\xi$.
 
Finally, taking advantage of the NI-based machinery developed here, we go one step further, and study the $\xi$-dependence of the gluon two-point function, which, in the low momentum region under scrutiny translates directly  into a statement on the dynamically generated gluon mass. The relevant auxiliary functions are  evaluated using again the approximations and assumptions employed in the previous case. The result reveals that the $\xi$-derivative of the gluon mass displays an IR logarithmic divergence, which can be traced back to the masslessness of the ghost propagator. As we explain in terms of an explicit example, such a divergent derivative may originate from perfectly IR finite gluon propagators, such as those found in the lattice simulations of~\cite{Cucchieri:2011pp} for $\xi\ll1$.

The article is organized as follows. In Sect.~\ref{SDE} we set up the $R_\xi$ ghost gap equation, discuss the approximations and assumptions employed, and present its numerical solutions,  paying particular attention to the deep IR behavior. In Sect.~\ref{NI} we address the same problem from the point of view of the NIs. Focusing on the identity satisfied by the ghost dressing function, we evaluate it  numerically within the one-loop dressed approximation, which allows for the determination of the leading IR behavior of $F$. The result turns out to be in excellent qualitative agreement with that found in the previous section. In Sect.~\ref{sec:gluon-2p} the NI analysis is extended to the gluon propagator. In particular,  a constraint on the IR behavior of the dynamical gluon mass is obtained, and an Ansatz for the possible $\xi$-dependence of the gluon mass is proposed. Our conclusions are presented in Sect.~\ref{sec:concl}. Finally, the technical details necessary to derive the Yang-Mills NIs are summarized in Appendix~\ref{app:Nielsen}.

\section{\label{SDE}Schwinger-Dyson equation analysis}

In this section we carry out a general analysis of the SDE that governs the ghost propagator, and eventually its dressing function.

\subsection{\label{SDEapprox}General considerations and approximations}
\begin{figure}[!t]
\centerline{\includegraphics[scale=0.65]{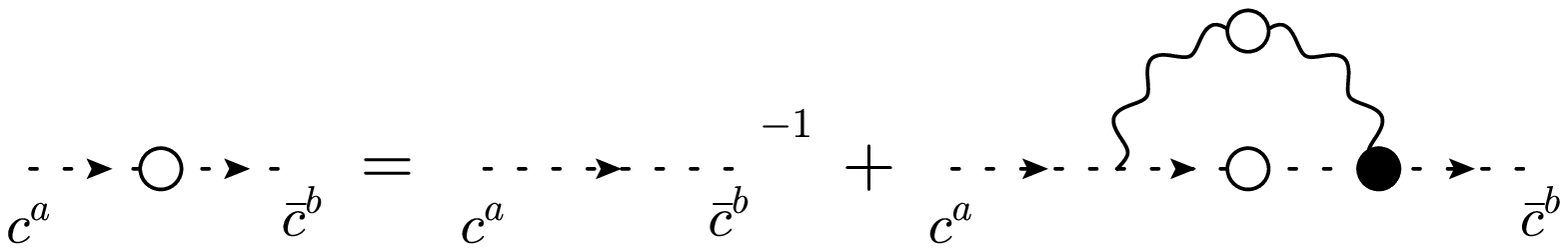}}
\caption{\label{fig:ghost-SDE}The ghost gap equation. White (respectively, black) blobs represent connected (respectively, one-particle irreducible) Green's functions.}
\end{figure}

The ghost gap equation (\fig{fig:ghost-SDE}) can be obtained directly from the one-loop ghost self-energy equation by fully dressing the internal gluon and ghost lines and one of the gluon ghost vertices appearing in it~\cite{Roberts:1994dr}. Dressing the right vertex, the SDE for the ghost propagator in a linear covariant gauge reads (factoring out the trivial color structure $\delta^{ab}$)
\begin{align}
\Dxi^{-1}(q^2)&=q^2-i\Pi(q^2)\nonumber \\
&=q^2+i\gA\int_k(k+q)^\mu \Dxi(k+q)\Delta_{\mu\nu}(k)\Gamma^\nu(k+q,-k,-q),
\label{ghSDE-rightdr}
\end{align}
where $\Pi(q^2)$ represent the ghost self-energy, $C_A$ is the Casimir eigenvalue of the adjoint representation, and the integral measure is defined as 
~\mbox{$\int_{k}\equiv \mu^{\epsilon}/(2\pi)^{d}\!\int\!\mathrm{d}^d k$}, with $\mu$ the 't Hooft mass and $d=4-\epsilon$ the dimension of the space-time.
$\Delta_{\mu\nu}$ and $D$ denote, respectively, the $R_\xi$ gluon and ghost propagators, defined according to\footnote{Our conventions can be found in Appendix~\ref{app:Nielsen}.} 
\begin{align}
i\Delta_{\mu\nu}(q)&=-i\left[P_{\mu\nu}(q)\Deltaxi(q^2)+\xi\frac{q_\mu q_\nu}{q^4}\right];& P_{\mu\nu}(q)=g_{\mu\nu}-\frac{q_\mu q_\nu}{q^2},\nonumber\\
iD(q^2)&=i\frac{F(q^2)}{q^2},
\label{defprop}
\end{align}
where $\xi$ is the non-negative gfp~\cite{Fujikawa:1972fe} (see also Appendix~\ref{A1}), and $F(q^2)$ is the so-called ghost ``dressing function''.
$\Gamma^\nu$ represents the full ghost-gluon vertex, with (all momenta entering)
\be
\Gamma^\nu(q_1+q_2,-q_1,-q_2) ={\cal A}(q_1+q_2,-q_1,-q_2)q_2^\nu+{\cal B}(q_1+q_2,-q_1,-q_2)q_1^\nu,
\label{defvert}
\ee
where $q_1$ ($q_2$) is the gluon (antighost) momentum;
at tree-level, ${\cal A}^{(0)}=1$ and ${\cal B}^{(0)}=0$. 

Then, using~\2eqs{defprop}{defvert}, we may rewrite~\1eq{ghSDE-rightdr} as
\begin{align}
\Dxi^{-1}(q^2)&=q^2+i\gA q^\mu q^\nu\int_k\Dxi(k+q)\Deltaxi(k)P_{\mu\nu}(k){\cal A}\nonumber \\
&+i\xi\gA\int_k  \Dxi(k+q) \left(1+\frac{k\cd q}{k^2}\right) \left({\cal B}+\frac{k\cd q}{k^2} {\cal A} \right),
\label{ghSDE-rightdr-1}
\end{align}
where the common argument $(k+q,-k,-q)$ of the form factors  ${\cal A}$  and ${\cal B}$ 
has been suppressed.

Solving this  equation in its  full generality would require  either independent knowledge 
of  the gluon propagator  and the form factors of the ghost vertex  for  general $\xi$,  or to couple 
\noeq{ghSDE-rightdr-1}  to   the  corresponding   SDEs  describing
$\Delta$, ${\cal A}$  and ${\cal B}$. However, apart
from  the  lattice   study of~\cite{Cucchieri:2011pp}, which  investigated 
the gluon propagator for very small values of $\xi$ ($\xi< 10^{-3})$ , there
is no direct knowledge of  the aforementioned quantities. As for
solving  the  full  coupled  system  of SDEs, unfortunately it constitutes a task that lies beyond our present powers. 

Therefore, we will instead study the SDE of~\1eq{ghSDE-rightdr-1} within the one-loop dressed approximation, 
which is obtained by keeping the propagators fully dressed and assigning   
tree-level values to ${\cal A}$ and ${\cal B}$. In addition, 
we will approximate the $\Delta(q^2)$ appearing in the first term on the right-hand side (rhs) of~\1eq{ghSDE-rightdr-1}
by the Landau gauge propagator $\DeltaL(q^2)$. 
The main underlying assumptions behind this later approximation are that $\Delta(q^2)$ saturates in the IR, assuming the standard form
\begin{equation}
\Delta^{-1}(q^2)=q^2J(q^2)-m^2(q^2),
\label{J-m}
\end{equation}
and that the deviation between $\Delta^{-1}(q^2)$ and  $\DeltaL(q^2)$ in the 
intermediate momenta region is relatively mild, at least for $0\le\xi\le 1$. 
Of course, as one approaches the region of larger momenta, the perturbative 
behavior will eventually set in; at one-loop order, $\Delta^{-1}(q^2)$ renormalized in the momentum-subtraction (MOM) scheme  
is given by 
\be
\Delta^{-1}(q^2) \sim q^2J(q^2) = q^2 \left[1+ \frac{\alpha_s C_A}{8\pi} \left(\frac{13}{3} - \xi \right)\log \frac{q^2}{\mu^2}\right ],
\ee
where $\mu$ is the renormalization point. For example, for the typical values used in this work, \ie  $\mu=4.3$ and $\alpha(\mu^2)=0.22$, the difference between the Landau and Feynman gauge perturbative tails is no more than 7\% in the momenta range $2\div 5$ GeV. 
In any case, as will become clear in the ensuing analysis, the behavior of $F(q^2)$ in the deep IR 
is not particularly sensitive to the above considerations; in fact, 
the complete knowledge of the gluon propagator would only affect the subleading terms.

Thus, the simplified version \noeq{ghSDE-rightdr-1} that we will consider is given by 
\be
\Dxi^{-1}(q^2) = q^2+i\gA \int_k\Dxi(k+q) \left[q^\mu q^\nu\DeltaL(k)P_{\mu\nu}(k) 
+\xi \left(1+\frac{k\cd q}{k^2}\right)\frac{k\cd q}{k^2} \right].
\label{ghSDE-simple0}
\ee
This particular integral equation must be properly renormalized, through the introduction of the appropriate renormalization constants for $\Dxi$, $\DeltaL$, and $\xi$. As is well-known, in principle the complete renormalization procedure must be carried out multiplicatively. As a result, in addition to the ghost renormalization constant $Z_c$ that will multiply the tree level term  $q^2$, further constants multiplying the remaining terms on the rhs of \1eq{ghSDE-simple0} must be included; this, in turn, adds an inordinate amount of complexity to the entire problem. Following the standard approximation, we will simply replace $q^2 \to Z_c q^2$, and set all multiplicative constants equal to unity, thus employing subtractive instead of multiplicative renormalization~\cite{Curtis:1990zs,Kizilersu:2009kg}. The actual expression for $Z_c$ is fixed from \1eq{ghSDE-simple0} through the momentum subtraction (MOM) renormalization condition $\Dxi^{-1}_\s{\rm R}(\mu^2_\s{\rm R}) = \mu^2_\s{\rm R}$, where $\mu^2_\s{\rm R}$ is the renormalization point. 

As an elementary check, we may recover from \1eq{ghSDE-simple0} the one-loop expression for $F(q^2)$. In particular, setting tree-level values for $\Dxi(k+q)$ and $\DeltaL(k)$, it is straightforward to show that
\be
F^{-1}(q^2) = 1 + \frac{i\gA}{4} \left[(3-\xi)\int_k \frac{1}{k^2(k+q)^2} + 2(1-\xi)\int_k  \frac{k\cd q}{k^4(k+q)^2} \right].
\label{Foneloop}
\ee
Using standard integration formulas, setting $\dE{q}^2=-q^2$, and renormalizing in the aforementioned scheme, one obtains for the renormalized ghost dressing function 
\be
F^{-1}_{\rm R}(\dE{q}^2) = 1 + \frac{\alpha_s C_A}{16\pi}(3-\xi) \log(\dE{q}^2/\mu^2),
\ee
where we have defined $\alpha_s=g^2/4\pi$.

\subsection{\label{SDEnum}Numerical analysis}

After a set of basic algebraic manipulations, together with the shift $k+q \to k$, we may cast \1eq{ghSDE-simple0} in the form
\be 
\Dxi^{-1}(q^2) = q^2+i\gA \int_k \Dxi(k) \left\{\frac{q^2k^2-(k\cd q)^2}{(k+q)^2}\DeltaL(k+q) 
+\frac{\xi}{4}\left[\frac{(k^2-q^2)^2}{(k+q)^4}-1 \right]\right\}.
\label{ghSDE-simple}
\ee
This last form of the ghost SDE is more convenient for the numerical analysis that follows, because it allows us to carry out exactly the angular integration in the term proportional to $\xi$, while in the first term the angular dependence has been passed from the unknown function $D(k+q)$ to the function $\DeltaL(k)$, which is known from the lattice.

In order to solve this equation, we first pass to Euclidean space using the~standard substitution rules 
\begin{align}
{\rm d}^4k &\to i{\rm d}^4\dE{k};&
(q^2,k^2,k\spr q)&\to (-\dE{q}^2,-\dE{k}^2,-\dE{k}\cdot\dE{q});&
\Delta (q^2), D(q^2) \to -\Delta_{\rm{E}}(\dE{q}^2), -D_{\rm{E}}(\dE{q}^2),
\end{align}
and suppress throughout the  subscript ``E'' in what follows.
Next, we introduce spherical coordinates (in $d=4$), through the relations
\begin{align}
& x=q^2;\qquad y=k^2;\qquad z=(k +q)^2= x+y+2 \sqrt{xy}\cos\theta;\nonumber\\
& \int_{k_\s{\mathrm{E}}}=\frac1{(2\pi)^3}\int_0^\pi\!\mathrm{d}\theta\,\sin^2\theta\int_0^\infty\!\mathrm{d}y\,y,
\label{sph-c}
\end{align}
use the result
\be
\int_0^\pi\!\diff\theta\,\frac{\sin^2\theta}{z^2} =\frac\pi2\left[\frac1{x(x-y)}\Theta(x-y)+\frac1{y(y-x)}\Theta(y-x)\right],
\ee
where $\Theta(x)$ is the Heaviside function, and factor out a $q^2$ from both sides of \1eq{ghSDE-simple}. Thus, we obtain the final equation  for the (subtractively renormalized) ghost dressing function $F(x)$,
\begin{align}
F^{-1}(x)&= Z_c - \frac{\alpha_s C_A}{2\pi^2} \int_0^\infty\!\diff y\,y\,F(y) \int_0^\pi\!\diff\theta\,\frac{\sin^4\theta}z\DeltaL(z)\nonumber \\
&+\xi\frac{\alpha_s C_A}{16\pi}\left[\frac1{x^2}\int_0^x\!\diff y\,yF(y)+\int_x^{\infty}\!\diff y\,\frac{F(y)}y\right],
\label{ghSDE-final}
\end{align}

Before proceeding to the full numerical treatment of this integral equation, it would be useful to identify some of its main IR features by means of a more direct method. In particular, if we assume that the $F(x)$ reaches a finite value in the IR ($x \to 0$), inspection of \1eq{ghSDE-final} reveals that the dominant term in that momentum region is the last one. Indeed, the first term corresponds {\it qualitatively} to the Landau gauge case: if the gluon propagator ($\DeltaL$) saturates in the IR, this term is finite. The second term is also finite in the IR, as the simple change of variable $y=t x$ immediately demonstrates. Therefore, keeping only the dominant IR contribution on the rhs of 
\1eq{ghSDE-final} we obtain 
\begin{equation}
F^{-1}(x)\underset{x\to0}{\sim} \xi\frac{\alpha_s C_A}{16\pi}\int_x^{\infty}\!\diff y\,\frac{F(y)}y.
\end{equation}

This integral equation can be converted into a differential equation, 
by differentiating both sides with respect to $x$; we then obtain
\begin{equation}
F'(x)\underset{x\to0}{\sim} \xi \c\frac{F^3(x)}x;\qquad \c=\frac{\alpha_s C_A}{16\pi},
\end{equation}
which is solved by
\begin{equation}
F(x)\underset{x\to0}{\sim}\pm\frac1{\sqrt{\a-2\xi \c\log(x/\mu^2)}},
\label{IR-sol}
\end{equation}
with $\a$ a (possibly $\xi$ dependent) constant, and $\mu$ a suitable renormalization scale; 
the physical solution corresponds to the positive sign. Notice that  
the IR solution given in~\1eq{IR-sol}
requires the aforementioned non-negativity condition $\xi\ge0$, since otherwise $F$ would become complex; in particular,
from now on, we will restrict our attention to $\xi\in[0,1]$.

\1eq{IR-sol} predicts an important qualitative modification  
in the IR behavior of the ghost dressing functions, compared to what is known from the Landau gauge studies.
Specifically,  whereas in the Landau gauge $\FL(0)=$const, whenever $\xi>0$ one finds 
that $F$ is driven to zero at the origin, namely $F(0)=0$.

\begin{figure}[!t]
\centerline{\includegraphics[scale=0.725]{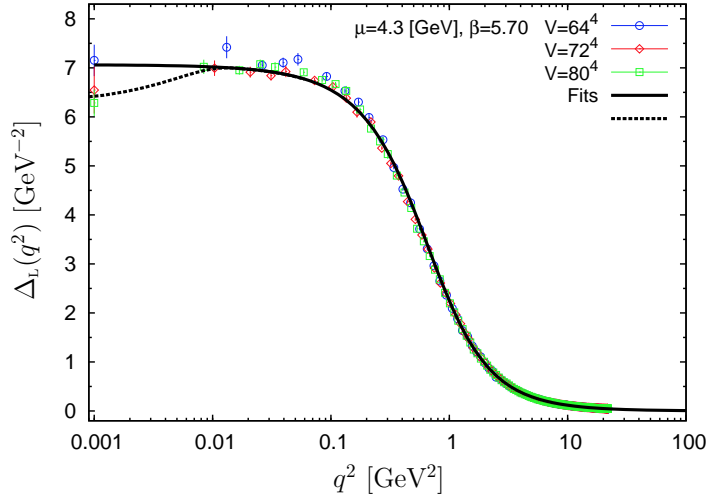}}
\caption{\label{fig:lattice-gluon-input} (color online). The lattice SU(3) gluon propagator evaluated in the Landau gauge~\cite{Bogolubsky:2009dc} and the corresponding fit used in our calculation~\cite{Aguilar:2010gm}. The dashed curve shows a fit featuring an IR maximum which is due to the presence of (divergent) contributions to the gluon (inverse) dressing function~\cite{Aguilar:2013vaa}. All functions are renormalized at~$\mu=4.3$~GeV.}  
\end{figure}

We next focus on the complete numerical evaluation of \1eq{ghSDE-final}. To this end, we will
use as input for $\DeltaL$
the fit to the available SU(3) lattice data~\cite{Bogolubsky:2009dc} introduced in~\cite{Aguilar:2010gm} 
(see \fig{fig:lattice-gluon-input}). The value of the renormalization point within the MOM scheme is $\mu=4.3$ GeV. Notice that in~\fig{fig:lattice-gluon-input} we show also a fit displaying an IR maximum that must appear due to the presence of divergent terms contributing to the gluon (inverse) dressing function~\cite{Aguilar:2013vaa} (see also Sect.~\ref{sec:smallxi}); however, the results finally obtained from the solution of the SDE are completely insensitive to the implementation of this particular feature in the gluon propagator.

\begin{figure}[!t]
\centerline{\mbox{}\hspace{-1.5cm}\includegraphics[scale=0.75]{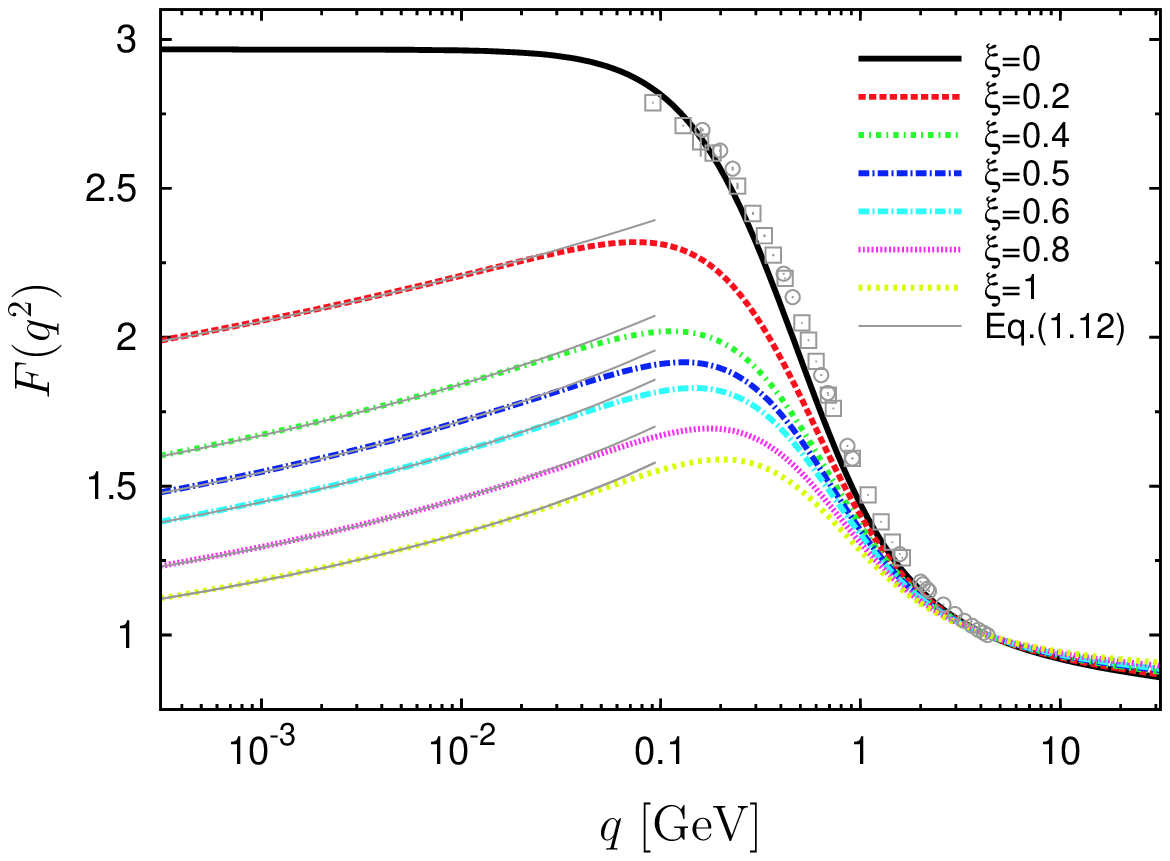}}\vspace{0.5cm}
\centerline{\mbox{}\hspace{-1.5cm}\includegraphics[scale=0.75]{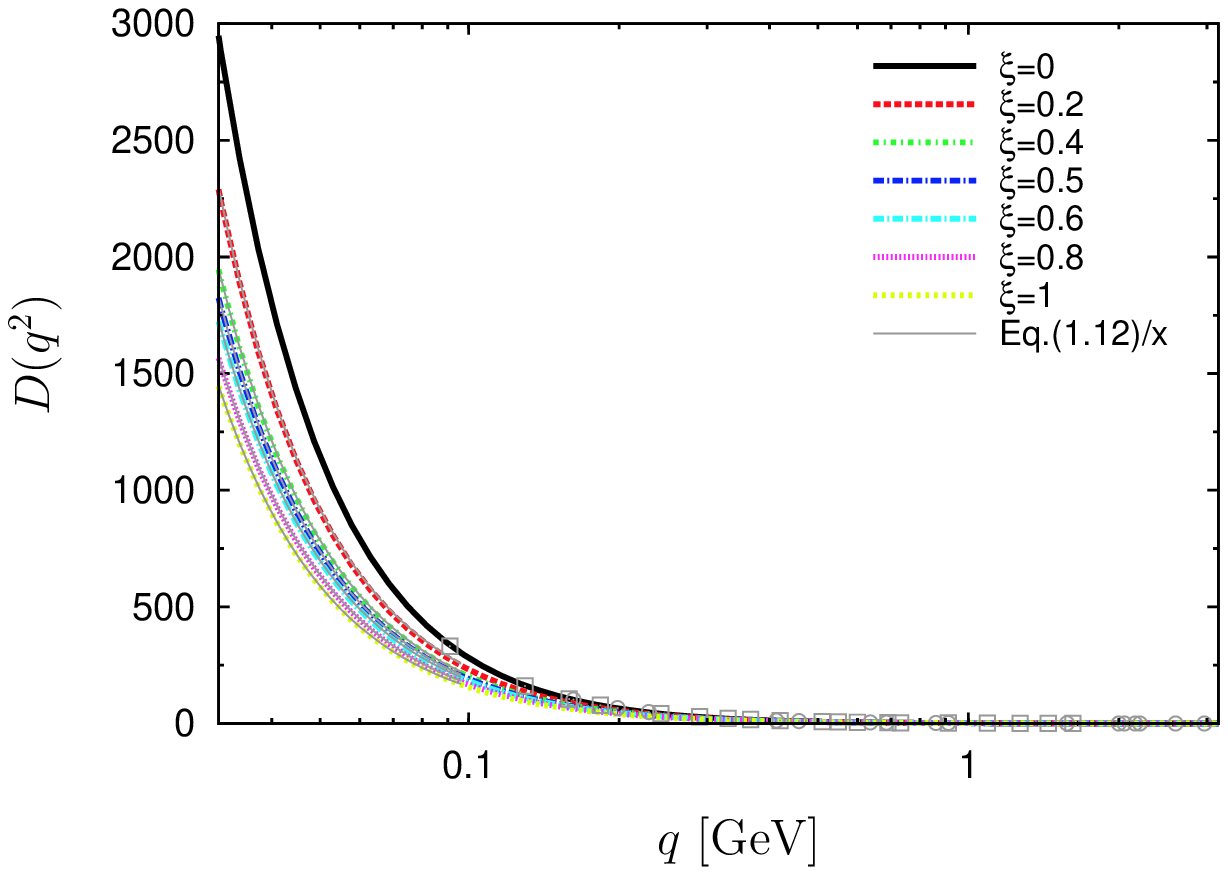}}
\caption{\label{fig:SDE-various-xi} (color online). Solution of the SDE~\noeq{ghSDE-final} (upper panel) and the associated ghost propagator (lower panel) for various values of the gauge fixing parameter~$\xi$. In the IR the solution obtained is perfectly described by~\1eq{IR-sol} after fitting for determining the value of the arbitrary constant $a$. For comparison we plot also the Landau gauge lattice data of~\cite{Bogolubsky:2009dc}.}  
\end{figure}

The solutions obtained for $\alpha_s=0.29$ and gfp values  ranging from 0 to 1 are shown in the left panel of~\fig{fig:SDE-various-xi}. The value of $\alpha_s$ is chosen so that in the Landau gauge $\xi=0$ one reproduces the lattice data of~\cite{Bogolubsky:2009dc} (see the black continuous curve  in~\fig{fig:SDE-various-xi}); the 30\% deviation from the expected value of $\alpha_s=0.22$  (at $\mu=4.3$ GeV) is due to the use of the tree-level ghost-vertex, as demonstrated in~\cite{Aguilar:2013xqa}.

One immediately observes the drastic change in the IR behavior of the ghost dressing function: at $\xi=0$ $\FL(0)$ is finite, whereas when $\xi\neq0$ $F(0)$ vanishes. The IR behavior is precisely the one described by the IR solution~\noeq{IR-sol}, where
\begin{equation}
\a=\a(\xi)=0.12(1+\xi); \qquad \c=0.035.
\end{equation}
Evidently, the rate at which $F(q^2)$ approaches zero is very slow, and begins to set 
on at the rather low scale of about $100$ MeV  
(upper panel of \fig{fig:SDE-various-xi}). However, the first appreciable deviations from the 
$\FL(q^2)$ obtained in the Landau gauge manifest themselves at the higher scale of about $300$ MeV, where the $F(q^2)$ displays 
a characteristic maximum. This particular feature, in turn, may serve as a guiding signal in  
future lattice simulations away from the Landau gauge.
 
The overall effect of $F(q^2)$ on the full ghost propagator $D(q^2)$ is shown in the right panel of \fig{fig:SDE-various-xi}. In particular, one observes that the rate of divergence of the ghost propagator at the origin becomes slightly softer compared to that of the Landau gauge.

Let us conclude this section by determining for later convenience  the IR behavior 
of the derivative with respect to $\xi$ of the ghost dressing evaluated at $\xi=0$; one finds
\begin{align}
\left.\partial_\xi F(x)\right\vert_{\xi=0}&\underset{x\to0}{\sim}\cSDE\log\frac x{\mu^2}\times\FL(0);&
\cSDE&=\frac{\alpha_s C_A}{16\pi}\frac{1}{a(0)},
\label{dbydxiSDE}
\end{align}   
where we have used the fact that $\FL(0)=1/\sqrt{a(0)}$. 
Clearly, this quantity displays an IR logarithmic divergence; substituting the numerical values of the constants involved one obtains~$\cSDE=0.15$.

\section{\label{NI}Nielsen identities}

In this section we take a different but complementary look at the problem, by resorting to 
a set of identities originally introduced by Nielsen~\cite{Nielsen:1975fs,Nielsen:1975ph}; 
for all technical details the reader is referred to Appendix~\ref{app:Nielsen}, where the general derivation is summarized. 

\subsection{\label{NIghost1}Ghost propagator}

Consider the ghost two-point sector of the theory. The corresponding  NI is readily obtained by differentiating the functional identity~\noeq{NId-final} with respect to one antighost and one ghost field; setting  afterwards all fields and sources to zero, one obtains the relation
\begin{equation}
\partial_\xi\Gamma_{c^a\bar c^b}(q^2)=i\Gamma_{\bar c^b\chi A^d_\mu}(q,0,-q)\Gamma_{c^a A^{*\mu}_d}(q)-i\Gamma_{c^a\chi c^*_d}(q,0,-q)\Gamma_{c^d\bar c^b}(q^2),
\label{gh-Nielsen}
\end{equation}
where 
\begin{equation}
\Gamma_{c^a\bar c^b}(q^2)=-i\delta^{ab}q^2F^{-1}(q^2);\qquad 
\Gamma_{c\bar c}(q^2)=-\Pi(q^2).
\label{twop-F}
\end{equation}
In~\1eq{gh-Nielsen} $\phi^*$ denotes the antifield associated to the field $\phi$. In addition, $\chi$ represents the static (\ie momentum independent) source associated to the gfp $\xi$; therefore, and despite their appearance, all functions in the identity above are two-point functions.

\1eq{gh-Nielsen} can be further simplified by noticing that the so-called ghost (or Faddeev-Popov) equation~\noeq{FPEq} yields
\begin{equation}
\Gamma_{c^aA^{*b}_\mu}(q)=i\delta^{ab}\frac{q_\mu}{q^2}\Gamma_{c^a\bar c^b}(q^2),
\end{equation}
a result which allows to trade the function $\Gamma_{cA^*}$ in~\noeq{gh-Nielsen} for a ghost two-point function $\Gamma_{c\bar c}$. Then, factoring out the trivial color structure $\delta^{ab}$, one is left with the identity
\begin{align}
\partial_\xi\Gamma_{c\bar c}(q^2)&=-\left[\frac{q^\mu}{q^2}\Gamma_{\bar c\chi A_\mu}(q,0,-q)+i\Gamma_{c\chi c^*}(q,0,-q)\right]\Gamma_{c\bar c}(q^2).
\label{NId-last}
\end{align}

\begin{figure}
\centerline{\includegraphics[scale=0.85]{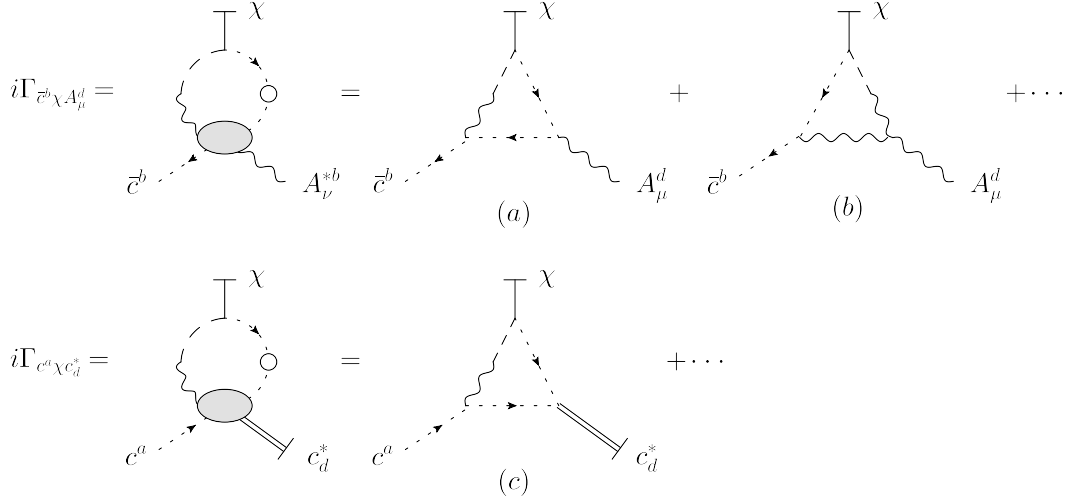}}
\caption{\label{fig:ghost-NIs-diagrams}One-loop diagrams contributing to the auxiliary functions  $\Gamma_{\bar c\chi A_\mu}$ and $\Gamma_{c\chi c^*}$ appearing in the ghost two-point Nielsen identity~\noeq{gh-Nielsen}. Notice the presence of the mixed propagator $\Delta_{bA}$.}
\end{figure}

In order to appreciate with a concrete example how the NIs work, 
let us consider the explicit realization of~\1eq{NId-last} at the one-loop level. 
The left-hand side (lhs) of \1eq{NId-last}
can be immediately deduced from~\1eq{Foneloop}, yielding 
\begin{equation}
\partial_\xi\Gamma^{(1)}_{c\bar c}(q^2) = - \frac{g^2C_A}4 q^2\left[\int_k\frac1{k^2(k+q)^2}+ 2\int_k\frac{k\spr q}{k^4(k+q)^2}\right].
\label{SDE-result}
\end{equation}
Turning to the rhs of ~\noeq{NId-last},  the diagrams contributing to the  auxiliary functions  $\Gamma_{\bar c\chi A_\mu}$ and $\Gamma_{c\chi c^*}$ at one-loop level are shown in~\fig{fig:ghost-NIs-diagrams}. Using the Feynman rules reported in Appendix~\ref{app:Nielsen} and Ref.~\cite{Binosi:2008qk}, one 
has the results
\begin{align}
i\Gamma^{(1)}_{\bar c\chi A_\mu}(q,0,-q)&=\frac{g^2C_A}2\left[q_\sigma\int_k\frac1{k^4}P^\sigma_\mu(k+q)-\int_k\frac{k\spr q}{k^4(k+q)^2}(k+q)_\mu\right],
\nonumber \\
i\Gamma^{(1)}_{c\chi c^*}(q,0,-q)&=i\frac{g^2C_A}2\int_k\frac{k^2+k\spr q}{k^4(k+q)^2}.
\end{align}
Notice that the contribution proportional to $\xi$
that could be in principle generated from diagram $(b)$ of~\fig{fig:ghost-NIs-diagrams} vanishes as a result of the Slavnov-Taylor identity \mbox{$q_1^\mu q_2^\nu q_3^\rho\Gamma_{\mu\nu\rho}(q_1,q_2,q_3)=0$}.
Thus one finally has
\be
\left[\frac{q^\mu}{q^2}\Gamma^{(1)}_{\bar c\chi A_\mu}(q,0,-q)+i\Gamma^{(1)}_{c\chi c^*}(q,0,-q)\right]\Gamma^{(0)}_{c\bar c}(q^2)
=\frac{g^2C_A}4 q^2\left[\int_k\frac1{k^2(k+q)^2}+ 2\int_k\frac{k\spr q}{k^4(k+q)^2}\right],
\ee
which, in view of \1eq{SDE-result}, confirms the validity of~\1eq{NId-last}  at one-loop.

\subsection{\label{sec:smallxi}Small $\xi$ limit and the one-loop dressed approximation}

Consider now the limit $\xi\ll1$; in this case, one can set $\xi=0$ on both sides of~\1eq{NId-last}, 
and use~\1eq{twop-F} to obtain
\begin{equation}
\left.\partial_\xi F(q^2)\right|_{\xi=0}=-\left[\frac{q^\mu}{q^2}\GammaL_{\bar c\chi A_\mu}(q,0,-q)+i\GammaL_{c\chi c^*}(q,0,-q)\right]\FL(q^2),
\label{NId-last-1}
\end{equation}
where the auxiliary ghost functions appearing on the rhs are now evaluated in the Landau gauge (see also the discussion at the end of Appendix~\ref{A1}).
This last equation can be used to deduce the IR behavior of $F(q^2,\xi)$  from the knowledge of the 
basic Green's functions in the Landau gauge. 
In particular, it allows us 
to compare the result obtained from the direct evaluation 
of the rhs of \1eq{NId-last-1} in the limit $q^2\to0$
with the corresponding expression derived in \1eq{dbydxiSDE} in the SDE context.  
To this end, we will study the auxiliary functions  $\GammaL_{\bar c\chi A_\mu}$ and $\GammaL_{c\chi c^*}$ 
in the one-loop dressed approximation, in which the diagrams contributing to each function 
are obtained from those shown in~\fig{fig:ghost-NIs-diagrams} by fully dressing the propagators, 
while keeping all vertices at their tree-level values\footnote{Note that the $b$-equation~\noeq{b-eq} implies that every Green's function which involves the 
Nakanishy-Lautrup multiplier $b$ remains fixed at its tree-level value: 
therefore in the $b$-sector the one-loop dressed approximation is exact.}.
The simple inspection of the diagrams given in~\fig{fig:ghost-NIs-diagrams} suggests that, indeed,  
a logarithmic behavior similar to that of \1eq{NId-last-1} is expected to make its appearance. 
This is because diagrams ($a$) and ($c$) 
may be essentially regarded as closed ghost loops, which, due to the 
nonperturbative masslessness of the ghost propagators entering in them, 
are known to diverge logarithmically in the IR~\cite{Aguilar:2013vaa,Tissier:2011ey}.

Let us then evaluate explicitly the one-loop dressed expressions of $\GammaL_{\bar c\chi A_\mu}$ and $\GammaL_{c\chi c^*}$;   
one has the following results
\begin{align}
\frac{q^\mu}{q^2}\GammaL_{\bar c\chi A_\mu}(q,0,-q)&\underset{\mathrm{1ldr}}{=}i\frac{g^2C_A}2\left[\int_k\frac{(k\spr q)(k\spr q+q^2)}{q^2k^4(k+q)^2}\FL(k)\FL(k+q)\right.\nonumber \\
&\left.-\int_k\frac{k^2q^2-(k\spr q)^2}{q^2k^4}\FL(k)\DeltaL(k+q)\right],\nonumber \\
i\GammaL_{\bar c \chi c^*}(q,0,-q)&\underset{\mathrm{1ldr}}{=}i\frac{g^2C_A}2\int_k\frac{k^2+k\spr q}{k^4(k+q)^2}\FL(k)\FL(k+q).
\label{1l-dr}
\end{align}
The terms proportional to the product of two ghost dressing functions $\FL$ in both functions are those corresponding to the aforementioned ghost-loops;
therefore, in the deep IR both functions display a logarithmic divergence, so that, in turn, one has  
\begin{equation}
\left.\partial_\xi F(q^2)\right\vert_{\xi=0}\underset{q^2\to0}{\sim} \cNI\log \frac{q^2}{\mu^2}\times\FL(0),
\label{IR-gh}
\end{equation}
where $\cNI$ a suitable constant and $\mu$ the renormalization scale chosen. 
Notice that this is exactly the kind of behavior found in~\1eq{dbydxiSDE} from the SDE analysis.

\begin{figure}[!t]
\centerline{\includegraphics[scale=0.75]{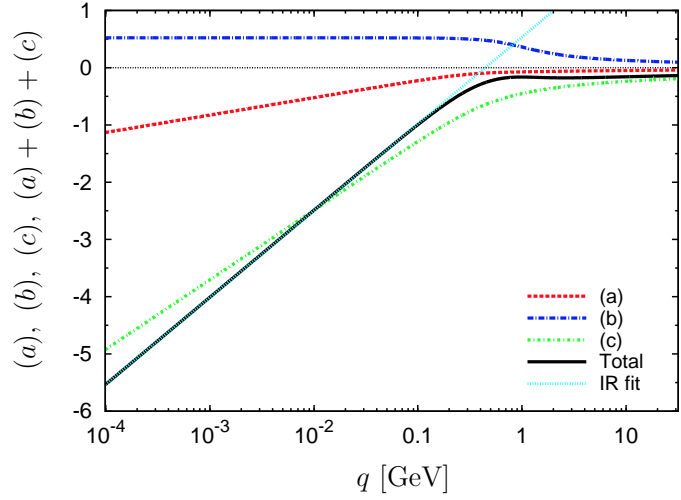}}
\caption{\label{fig:Ghost-NI} (color online). Contributions of the one-loop dressed auxiliary functions to the ghost two-point function Nielsen identity. The IR region is perfectly described by the predicted $\cNI\log q^2/\mu^2$ behavior yielding~$\cNI=0.33$.}  
\end{figure}

The qualitative agreement between \1eq{dbydxiSDE} and \1eq{IR-gh} motivates a further quantitative study, focusing 
on the actual value of the coefficient $\c$ obtained within the two methods (SDE vs NI).  
To accomplish this, 
we evaluate numerically the one-loop dressed contributions~\noeq{1l-dr}, which are given by (Euclidean space)
\begin{align}
\frac{q^\mu}{q^2}\GammaL_{\bar c\chi A}(q,0,-q)&\underset{\mathrm{1ldr}}{=}\frac{g^2 C_A}{2(2\pi)^3}\int_0^\pi\!\diff\theta\,\sin^2\theta\cos\theta\int_0^\infty\!\diff y\,\left(\cos\theta+\sqrt{\frac xy}\right)\frac1{z}\FL(y)\FL(z)\nonumber \\
&+\frac{g^2C_A}{3(2\pi)^3}\int_0^\pi\!\diff\theta\,\sin^4\theta\int_0^\infty\!\diff y\,\FL(y)\DeltaL(z)=(a)+(b),
\nonumber \\
i\GammaL_{\bar c \chi c^*}(q,0,-q)&\underset{\mathrm{1ldr}}{=}-\frac{g^2C_A}{2(2\pi)^3}\int_0^\pi\!\diff\theta\,\sin^2\theta\int_0^\infty\!\diff y\,\left(1+\sqrt{\frac xy}\cos\theta\right)\frac1{z}\FL(y)\FL(z)=(c),
\label{ints-final}
\end{align}
where $(a)$, $(b)$ and $(c)$ denote the contributions of the diagrams appearing in~\fig{fig:ghost-NIs-diagrams}. 
At this point all integrals can  be evaluated provided that we supply as input the Landau gauge gluon propagator $\DeltaL$ 
and the ghost dressing function $\FL$ (see~\fig{fig:lattice-gluon-input} and~\fig{fig:SDE-various-xi}, respectively).

The results obtained for the three individual terms $(a)$, $(b)$ and $(c)$ of~\1eq{ints-final}, as well as their sum, 
are shown on the left-panel of~\fig{fig:Ghost-NI}. One sees that  terms $(a)$ and $(c)$ show the claimed logarithmic divergence, 
while in the case of $(b)$ the gluon mass acts as an IR regulator, making the integral convergent. 
Adding the three contributions together one obtains the black continuous curve of~\fig{fig:Ghost-NI}, yielding the IR behavior~\noeq{IR-gh} 
with $\cNI=0.33$; this value should be compared to the value $\cSDE=0.15$ obtained from the SDE analysis.
Given that the two values are derived from two {\it a priori} completely distinct methods, we find the 
proximity between the two values rather encouraging.

\subsection{\label{sec:gluon-2p}Gluon propagator}

The NI formalism may be extended in a straightforward way to the case of the gluon propagator. Specifically,
the corresponding NI for the gluon two-point function $\Gamma_{AA}$
can be derived by differentiating~\1eq{NId-final} with respect to two gluon fields, and setting afterwards all fields to zero. In particular, one obtains the equation
\begin{equation}
\partial_\xi\Gamma_{A^a_\mu A^b_\nu}(q)=-i\Gamma_{A^a_\mu\chi A^{*\rho}_c}(q,0,-q)\Gamma_{A^c_\rho A^b_\nu}(q)-i\Gamma_{A^b_\nu\chi A^{*\rho}_c}(q,0,-q)\Gamma_{A^c_\rho A^a_\mu}(q).
\label{gl-Nielsen}
\end{equation}
Given that $\Gamma_{AA}$ is transverse to all orders, with its tree-level value given by~\mbox{$\Gamma^{(0)}_{A^a_\mu A^b_\nu}(q)=iq^2\delta^{ab}P_{\mu\nu}(q)$} (see Appendix~\ref{app:Nielsen}), this identity can be further simplified to read
\begin{equation}
\partial_\xi\Gamma_{A A}(q^2)=-2i\Gamma_{A\chi A^*}(q,0,-q)\Gamma_{AA}(q^2),
\label{gl-Nielsen-1}
\end{equation}
where the color structure has been factored out, and we have defined
\begin{equation}
\Gamma_{A\chi A^*}(q,0,-q)=\frac1{d-1}P^{\mu\nu}(q)\Gamma_{A_\mu\chi A^*_\nu}(q,0,-q).
\end{equation}

\begin{figure}
\centerline{\includegraphics[scale=0.85]{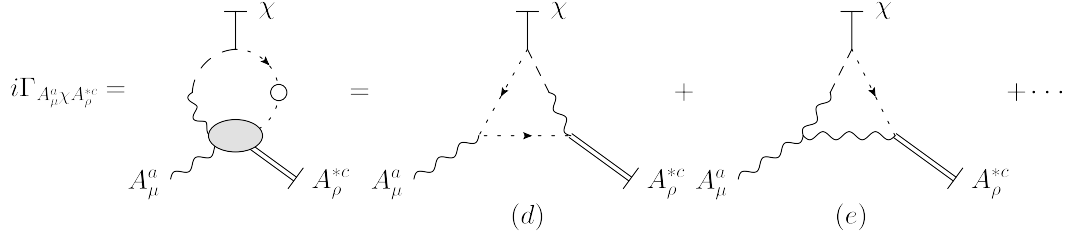}}
\caption{\label{fig:gluon-NIs-diagrams}One-loop diagrams contributing to the auxiliary function  $\Gamma_{A_\mu\chi A^*_\nu}$ appearing in the gluon two-point Nielsen identity~\noeq{gl-Nielsen}.}
\end{figure}

One can appreciate how the above identity works by evaluating it at lowest order in perturbation theory. 
The diagrams contributing to the function $\Gamma_{A\chi A^*}$ at the one-loop level are shown in~\fig{fig:gluon-NIs-diagrams}; then the rhs of~\1eq{gl-Nielsen-1} reads
\begin{align}
-2i\Gamma^{(1)}_{A_\mu\chi A^*_\rho}(q,0,-q)\Gamma^{(0)}_{A^\rho A_\nu}(q)&=g^2C_Aq^2P^\rho_\nu(q)\left\{\int_k\frac{(k^2-q^2)}{k^2(k+q)^2}P_{\mu\rho}(k)-\int_k\frac{(k+q)_\mu k_\rho}{k^4(k+q)^2}\right.\nonumber \\
&\left.+(1-\xi) q^2P_\mu^\sigma(q)\int_k\frac{k_\rho k_\sigma}{k^4(k+q)^4}\right\}.
\label{gl-aux}
\end{align}
To complete the comparison, note that $\Pi^{(1)}_{\mu\nu}(q)$ has been evaluated in~\cite{Binosi:2009qm} [see Eq.~(2.56)]; 
its derivative with respect to $\xi$  coincides with the result~\noeq{gl-aux}, 
once we take into account that \mbox{$\Pi_{\mu\nu}(q)=-\Gamma_{A_\mu A_\nu}(q)$}.

Next, we consider the $\xi\ll1$ limit of \1eq{gl-Nielsen-1}, obtaining
\begin{align}
\left.\partial_\xi\Delta^{-1}(q^2)\right|_{\xi=0}&=-2i\GammaL_{A\chi A^*}(q,0,-q)\DeltaL^{-1}(q^2);&
\Gamma_{AA}(q^2)=i\Delta^{-1}(q^2).
\label{NId-gl-1}
\end{align}
The rhs of this equation can then be evaluated within the one-loop dressed approximation, yielding the expression
\begin{align}
-2i\GammaL_{A\chi A^*}(q,0,-q)&\underset{\mathrm{1ldr}}{=}-i\frac{g^2C_A}{d-1}\left\{\int_k\frac{k^2q^2-(k\spr q)^2}{q^2k^4(k+q)^2}\FL(k)\FL(k+q)
\right.
\nonumber \\
&\left.+\int_k\frac{k^2-q^2}{(k+q)^4}\left[d-2+\frac{(k\spr q)^2}{k^2q^2}\right]\DeltaL(k)\FL(k+q)\right\}.
\end{align}
One notices again the presence of a massless ghost loop, which implies in turn the divergent IR behavior
\begin{equation}
\left.\partial_\xi\Delta^{-1}(q^2)\right\vert_{\xi=0}\underset{q^2\to0}{\sim}\left.\partial_\xi m^2(q^2)\right\vert_{\xi=0}\underset{q^2\to0}{\sim}\cNI\log \frac{q^2}{\mu^2}\times\m2L(0),
\label{IR-gl}
\end{equation}
where the first expression on the rhs originates from the fact that when $q^2\to0$, 
\mbox{$\Delta^{-1}(q^2) \to m^2(q^2)$} [see~\1eq{J-m}], and  
$\m2L(q^2)$ denotes the dynamical gluon mass in the Landau gauge~\cite{Aguilar:2011ux,Binosi:2012sj,Aguilar:2014tka}.

\begin{figure}[!t]
\centerline{\includegraphics[scale=0.75]{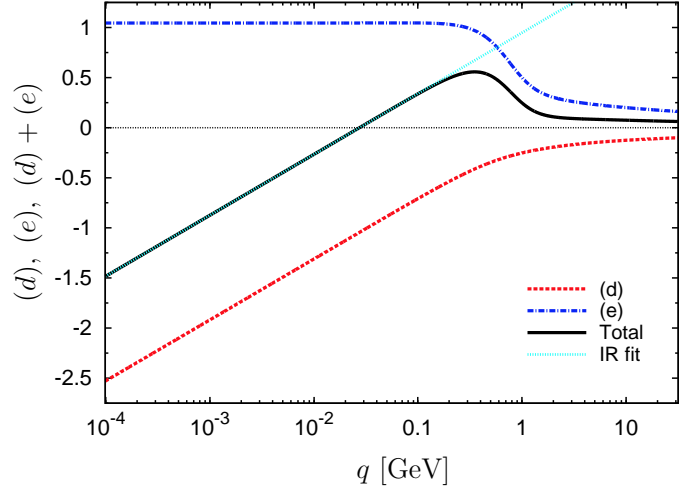}}
\caption{\label{fig:Gluon-NI} (color online). Contributions of the one-loop dressed auxiliary functions to the gluon two-point function Nielsen identity. Also in the gluon case the IR region is perfectly described by the predicted $b\log q^2/\mu^2$ behavior, now with~$\cNI=0.13$.}  
\end{figure}

The appearance of this particular behavior may be indeed confirmed numerically. Specifically,
after passing to the Euclidean metric and introducing spherical coordinates, one obtains 
\begin{align}
-2i\GammaL_{A\chi A^*}(q,0,-q)&\underset{\mathrm{1ldr}}{=}-\frac{g^2C_A}{3(2\pi)^3}\int_0^\pi\!\diff\theta\sin^4\theta\int_0^\infty\!\diff y\,\frac1{z_1}\FL(y)\FL(z_1)\nonumber \\
&+\frac{g^2C_A}{3(2\pi)^3}\int_0^\pi\!\diff\theta\,(3-\sin^2\theta)\sin^2\theta\int_0^\infty\!\diff y\,\frac{y(y-x)}{z_1^2}\DeltaL(y)\FL(z_1)\nonumber \\
&=(d)+(e),
\end{align}
which can be evaluated using the Landau gauge propagator and ghost dressing function introduced before.
The results are shown in~\fig{fig:Gluon-NI}; one observes a logarithmic IR divergence in diagram $(d)$, while the IR finiteness of diagram $(e)$ 
is due to the presence of the dynamical gluon mass. 
When summing everything together the IR behavior is indeed the one described by~\1eq{IR-gh}, with $\cNI=0.13$.

\begin{figure}[!t]
\mbox{}\hspace{-0.4cm}\includegraphics[scale=1]{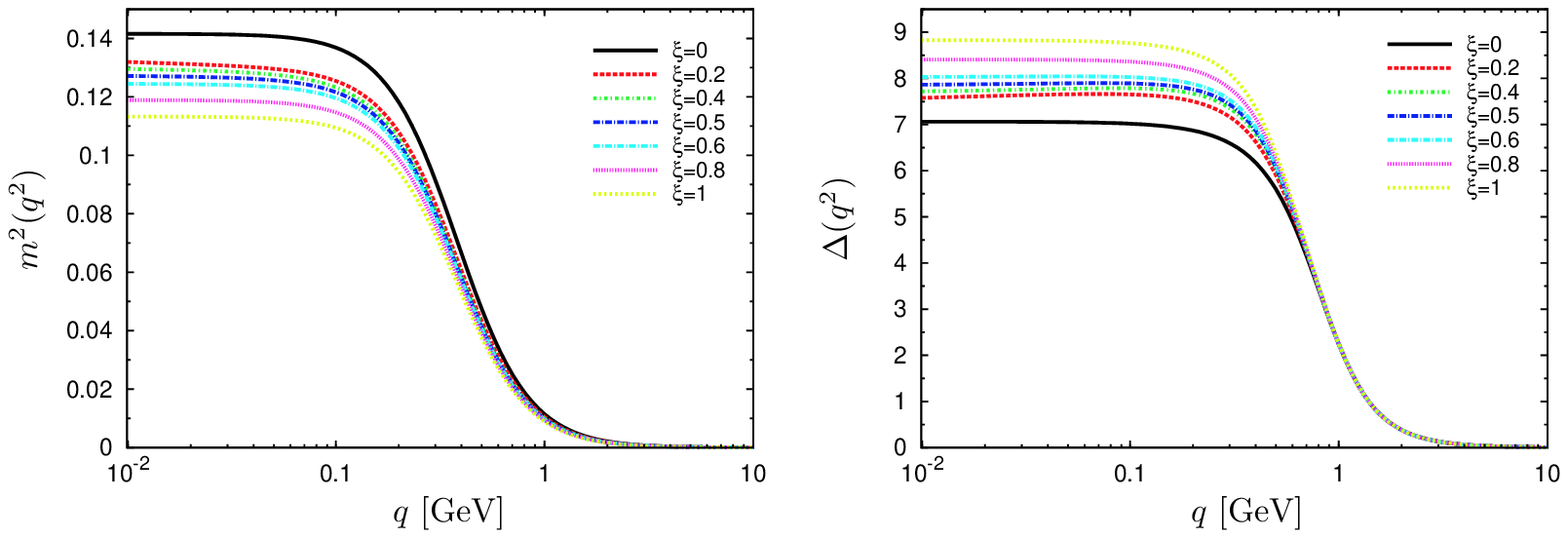}
\mbox{}\hspace{-0.4cm}\includegraphics[scale=1]{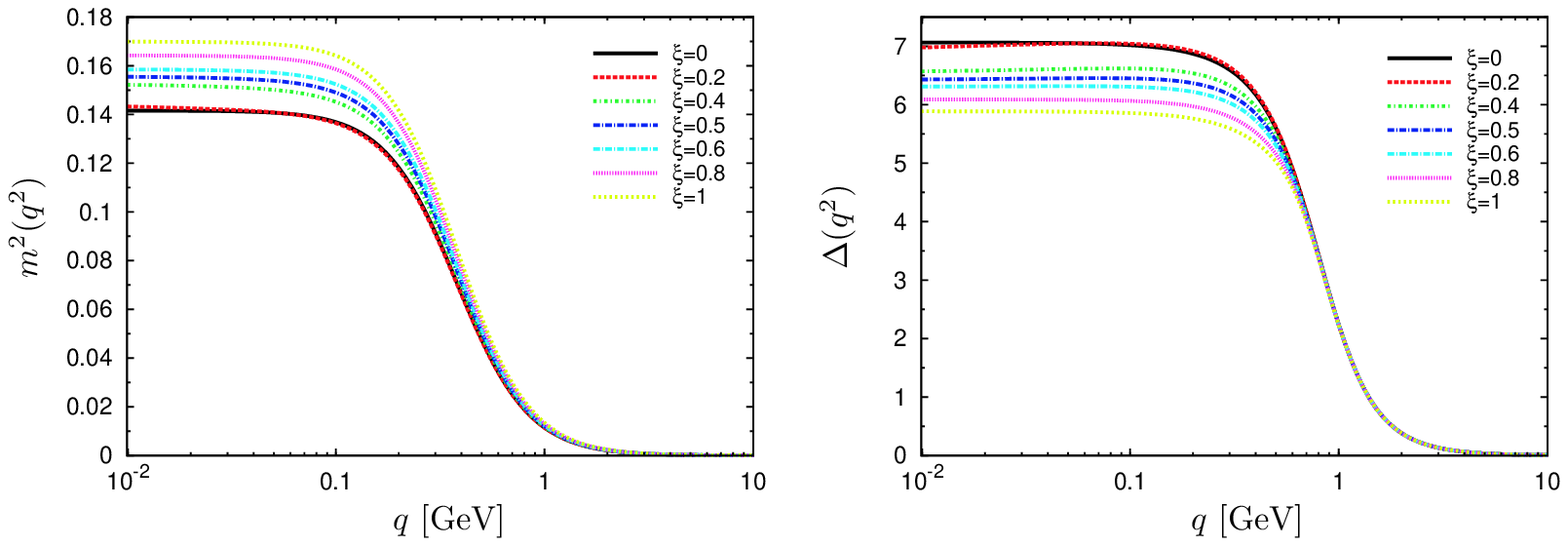}
\caption{\label{fig:aminus-aplus}(color online). $\xi$-dependence of the gluon mass (left panels) and gluon propagator (right panels) as predicted by the Ansatz~\noeq{mRxi} for $a_1=-0.2$ (upper panels) and $a_1=0.2$ (lower panels).}
\end{figure}

Finally, it is rather interesting to consider how  the IR divergence found in~\noeq{IR-gl} might be reconciled with the underlying assumption of an IR finite gluon propagator. Given that, at present, the dynamical equation that describes the gluon mass has only been derived in the Landau gauge\footnote{For related studies in the Coulomb gauge, see~\cite{Szczepaniak:2001rg,Szczepaniak:2003ve,Epple:2007ut,Szczepaniak:2010fe}}, one may only proceed by postulating an Ansatz for $m^2(q^2)$ that would satisfy ~\noeq{IR-gl}, and study its consequences at the level of the corresponding gluon propagators.

One such possibility is given by 
the following Ansatz for the $\xi$-dependent mass function\footnote{A simpler Ansatz would have been
$$m^2(q^2)=\left[\a(\xi)+\c(\xi)\left(\frac{q^2}{\mu^2}\right)^{\!\xi}\right]m^2_{\s{\mathrm L}}(q^2),$$
with $$\a(\xi)=1-\c_0-\a_1\xi+\cdots;\qquad \c(\xi)=\c_0+\cdots$$ and $\c_0\equiv \cNI\approx0.13$. In this case, however, 
the limits $\xi\to0$ and $q^2\to0$ do not commute, contrary to what happens with the  Ansatz~\noeq{mRxi}.}
\begin{equation}
m^2(q^2)=\left[\a(\xi)+\c(\xi)\left(\frac{q^2}{\mu^2}\right)^{\!\!\xi}\log\frac{q^2}{\mu^2}\right]m^2_{\s{\mathrm L}}(q^2),
\label{mRxi}
\end{equation}
with 
\begin{align}
\a(\xi)=\a_0+\a_1\xi+\cdots;\qquad \c(\xi)=\c_1\xi+\cdots.
\end{align}
Notice that the (resummed) behavior $\sim(q^2/\mu^2)^\xi$ has been also observed when studying the gfp-dependence of fermion propagators through LKF transformations~\cite{Curtis:1990zs,Bashir:2002sp}. 

Evidently, choosing $\a_0=1$ and $\c_1\equiv \cNI=0.13$ ensures that 
\begin{align}
m^2(q^2)&\underset{q^2\to0}{\sim}(1+\a_1\xi)\m2L(0),\nonumber \\
\left.\partial_\xi m^2(q^2)\right|_{\xi=0}&\underset{q^2\to0}{\sim}\c\log \frac{q^2}{\mu^2}\times \m2L(0),
\end{align}
in agreement with~\noeq{IR-gl}; in addition, small values of $a_1$ would make the $R_\xi$ and Landau-gauge  propagators and dynamical masses to be rather close to each other, justifying in retrospect  our replacing~$\Delta$ by ${\Delta}_{\s{\mathrm L}}$ when solving the ghost SDE.

Evidently, within this approach, the sign of the coefficient $a_1$ remains undetermined. This sign, in turn, controls the leading behavior of the gfp-dependence of the gluon mass (and correspondingly of the propagator) in the deep IR: a positive $a_1$ implies an increasing (decreasing) mass (propagator), while for $a_1$ negative the behavior would be reversed. This is shown in~\fig{fig:aminus-aplus}, where the left panels depict the $\xi$-dependence of the gluon dynamical mass~\noeq{mRxi} for the two values $a_1=0.2$ (upper-left) and $a_1=-0.2$ (lower left), while the corresponding gluon propagators, obtained from the relation $\Delta^{-1}(q^2)=q^2J_{\s{\mathrm L}}(q^2)+m^2(q^2)$, are shown on the right panels of the same figure.  

\begin{figure}[!t]
\includegraphics[scale=1]{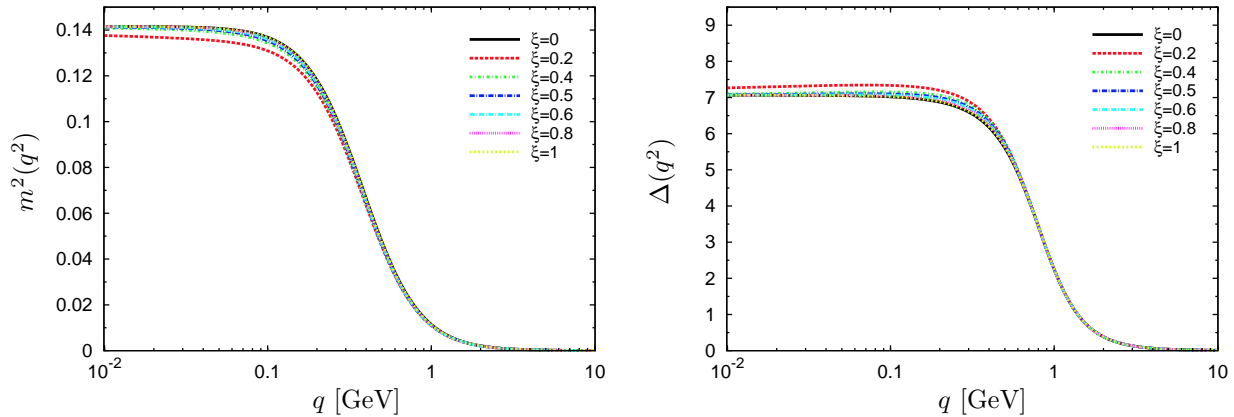}
\caption{\label{fig:a0}(color online). $\xi$-dependence of the gluon mass (left) and gluon propagator (right) as predicted by the Ansatz~\noeq{mRxi} for $a_1=0$.}
\end{figure}

Notice that the case $a_1=0$ would be particularly interesting, as it would imply that, at leading order in $\xi$, the $R_\xi$ gluon mass and propagator coincide in the IR with the corresponding quantities computed in the Landau gauge. In addition, as can be appreciated from~\fig{fig:a0}, the $\xi$-dependence over the entire range of momenta would be minimal.

\section{\label{sec:concl}Conclusions}

In the present work we have analyzed the nonperturbative behavior of Yang-Mills Green's functions quantized in a linear covariant gauge, paying particular attention to its dependence on the parameter $\xi$ characterizing this class of gauges. We have first focussed on the ghost two-point function and shown that, within a well-defined set of approximations, the solutions of the corresponding SDE for $\xi>0$ are such that the dressing function $F(q^2)$ vanishes as $q^2\to0$; this is in sharp contrast to the Landau gauge case ($\xi=0$) where $F(q^2)$ is known to saturate in the low momentum region. The particular IR behavior found for $F(q^2)$  turned out to be in notable agreement with that obtained from the Nielsen identity satisfied by this function, within the one-loop dressed approximation and for $\xi\ll1$. The NI analysis has been then extended to the gluon two-point function, and shown to predict the same kind of logarithmic divergence for the derivative with respect to $\xi$ of the dynamical gluon mass,  $\left.\partial_\xi m^2(q^2)\right\vert_{\xi=0}\sim\c\log q^2/\mu^2\times\m2L(0)$. A particular example of a $m^2(q^2)$ that reconciles this behavior with the assumed saturation of the gluon propagator away from the Landau gauge was given, and its main features were studied numerically.

Undoubtedly, lattice simulations would be crucial for verifying or amending the findings of this preliminary SDE study. As already mentioned, exploratory simulations in the linear gauges have already been carried out for the gluon propagator~\cite{Cucchieri:2011pp}; it would be interesting to extend them to larger values of $\xi$, in order to determine whether the observed IR saturation persists. Furthermore, the IR suppression of the ghost dressing function predicted here may serve as a definite reference when attempting to simulate the ghost sector of the theory. 

From the point of view of the SDEs, one may envisage various improvements. To begin with, the replacement of the fully-dressed ghost-gluon vertex by the tree-level expression inside the ghost SDE ought to be ameliorated. This, in turn, would require the treatment of the corresponding vertex SDE, for a general $\xi$, in the spirit of the analysis presented in the Landau gauge~\cite{Aguilar:2013xqa}. To be sure, subtractive instead of multiplicative renormalizability is another longstanding drawback in practically all types of SDE analysis; however, given that this problem cannot be even solved within the context of the (easier) Landau gauge, the prospects for a notable refinement in this particular direction seem rather reduced. 

It is also clear that additional theoretical work at the level of the gluon propagator is an absolute requirement before any firm statements could be made. In particular, no study related to the possibility of gluon mass generation away from the Landau gauge has been carried out to date; in the present work we have simply assumed the realization of this scenario, based almost  exclusively on the limited lattice evidence of~\cite{Cucchieri:2011pp}. In particular, it would be essential to derive the dynamical equation that governs the evolution of the gluon mass for an arbitrary $\xi$, and explore the type of solutions it might admit. This task is technically rather complex, mainly due to the proliferation of terms with respect to the Landau gauge case. Calculations in this direction are already in progress, and we hope to report progress in the near future.

\acknowledgments 

The research of J.~P. is supported by the Spanish MEYC under 
grant FPA2011-23596 and the Generalitat Valenciana under grant “PrometeoII/2014/066”. The work of  A.~C.~A  is supported by the 
National Council for Scientific and Technological Development - CNPq
under the grant 306537/2012-5 and project 473260/2012-3,
and by S\~ao Paulo Research Foundation - FAPESP through the project 2012/15643-1.

\appendix

\section{\label{app:Nielsen}Derivation of the Nielsen identities}

The action $\Gamma^{(0)}$ of the SU($N$) Yang-Mills theory can be written as the sum of three terms,
\begin{equation}
\Gamma^{(0)}=S_\s{\rm YM}+S_\s{\rm GF+FPG}+S_\s{\rm BV},
\label{action}
\end{equation}
where the first term corresponds to the classical action
\begin{align}
S_\s{\rm YM}&=-\frac14\int\!\diff^{4}x\,F^{a}_{\mu\nu}F_{a}^{\mu\nu};&
F^{a}_{\mu\nu}=\partial_\mu A^a_\nu-\partial_\mu A^a_\nu+g f^{abc}A^b_\mu A^c_\nu,
\end{align}
while the second to the gauge fixing and its associated Faddeev-Popov action, written as
\begin{equation}
S_\s{\rm GF+FPG}=s\int\!\diff^{4}x\,\bar c^a\left(
{\cal F}^a-\frac\xi2b^a
\right).
\end{equation}
In the equation above $b^a$ is the Nakanishy-Lautrup multiplier, $\bar c^a$ the antighost field, while ${\cal F}^a$ 
represents, for the moment, an arbitrary gauge fixing function. The only restriction on this latter function is that it allows for the inversion of the tree-level two-point functions of the $A$--$b$ sector, thus yielding the field propagators (in what follows we will use an off-shell formalism, keeping explicitly the $b$ fields, which, otherwise, can be eliminated by making use of their trivial equation of motion). Finally, $s$ is the BRST symmetry operator that acts on the elementary fields according to
\begin{align}
sA^a_\mu&=\underbrace{\left(\partial_\mu\delta^{ab}+gf^{acb}A^c_\mu\right)}_{{\cal D}^{ab}_\mu}c^b;&
sc^a&=-\frac12gf^{abc}c^bc^c;& 
s\bar c^a&=b^a;&
s b^a&=0,
\end{align}
with ${\cal D}^{ab}_\mu$ the usual covariant derivative.

As can be explicitly seen above, the BRST variations of the gauge and ghost fields are non-linear in the quantum fields; their renormalization is ensured by the introduction of external sources, known as antifields, in the third term of~\noeq{action}, reading
\begin{equation}
S_\s{\rm BV}=\int\,\diff^{4}x\,\left(
A^{*a}_\mu sA^\mu_a+c^*_asc^a
\right).
\end{equation}
The tree-level action~\noeq{action} will then satisfy the Slavnov-Taylor (ST) identity
\begin{align}
{\cal S}(\Gamma^{(0)})&=0;&
{\cal S}(\Gamma^{(0)})&=\int\!\diff^{4}x\,\left(\frac{\delta\Gamma^{(0)}}{\delta A^{*a}_\mu}\frac{\delta\Gamma^{(0)}}{\delta A_a^\mu}+\frac{\delta\Gamma^{(0)}}{\delta c^*_a}\frac{\delta\Gamma^{(0)}}{\delta c^a}+b^a\frac{\delta\Gamma^{(0)}}{\delta\bar c^a}\right).
\label{STid}
\end{align}
As the theory is anomaly-free, the ST identity~\noeq{STid} holds also for the full vertex functional~$\Gamma$.

If we extend the BRST to include also the gauge parameter $\xi$, we obtain an extended ST identity 
that gives control over the gfp-dependence of the Green's function of the theory~\cite{Nielsen:1975fs,Nielsen:1975ph}.
Writing\footnote{A pair of variables $(u,v)$ such that $su=v$ and $sv=0$ is called a BRST doublet; notice that also $\bar c$ and $b$ form such a doublet.}
\begin{equation}
s\xi=\chi;\qquad s\chi=0,
\end{equation}
one obtains that
\begin{equation}
S_\s{\rm GF+FPG}=\int\diff^{4}x\left(b^a{\cal F}^a
-\frac\xi2b_a^2-\bar c^as{\cal F}^a
\right)+\int\!\diff^{4}x\,\bar c^a\left(\frac12\chi b^a-\chi
\frac{\partial{\cal F}^a}{\partial \xi}\right),
\end{equation}
and therefore the tree-level action satisfies the extended ST identity
\begin{align}
{\cal S}'(\Gamma^{(0)})&=0;& {\cal S}'(\Gamma^{(0)})&={\cal S}(\Gamma^{(0)})+\chi\frac{\partial\Gamma^{(0)}}{\partial\xi}.
\end{align}
Again, the identity above is valid for the full vertex functional $\Gamma$; taking then a derivative with respect to $\chi$ and setting it to zero afterwards, one obtains the NI
\begin{align}
\left.\frac{\partial\Gamma}{\partial\xi}\right\vert_{\chi=0}&=\left.\int\!\diff^{4}x\,\left(
\frac{\delta\Gamma}{\delta A^{*a}_\mu}\frac{\delta^2\Gamma}{\partial\chi\delta A_a^\mu}
-\frac{\delta^2\Gamma}{\partial\chi\delta A^{*a}_\mu}\frac{\delta\Gamma}{\delta A_a^\mu}
-\frac{\delta^2\Gamma}{\partial\chi\delta c^*_a}\frac{\delta\Gamma}{\delta c^a}
-\frac{\delta\Gamma}{\delta c^*_a}\frac{\delta^2\Gamma}{\partial\chi\delta c^a}
-b^a\frac{\delta^2\Gamma}{\partial\chi\delta\bar c^a}\right)\right\vert_{\chi=0}.
\label{NId}
\end{align}

\subsection{\label{A1}Linear covariant gauges}

Even though the NI~\noeq{NId} holds irrespectively of the gauge fixing functional chosen, we will specialize from now on to the case of linear covariant (or $R_\xi$) gauges, which are identified by the choice of the following gauge fixing function
\begin{equation}
{\cal F}^a=\partial^\mu A^a_\mu,
\end{equation}
and, in our conventions, the non negativity condition on $\xi$~\cite{Fujikawa:1972fe}, needed to ensure that the (Euclidean) path integral over the $b$ fields is Gaussian. 

Thus in the two-point gluon sector the $R_\xi$ gauge fixing yields the tree-level propagators $AA$, $Ab$, $bb$ given respectively by
\begin{align}
i\Delta_{\mu\nu}^{ab}(q)&=-i\delta^{ab}\frac1{q^2}\left[P_{\mu\nu}(q)+\xi\frac{q_\mu q_\nu}{q^2}\right];& 
i\Delta^{ab}_\mu(q)&=\delta^{ab}\frac{q_\mu}{q^2};& 
i\Delta^{ab}&=0.
\end{align}
For the ghost sector the tree-level propagator is instead written as
\begin{equation}
iD^{ab}(q)=i\delta^{ab}\frac{1}{q^2}. 
\end{equation}

\begin{figure}
\centerline{\includegraphics[scale=0.85]{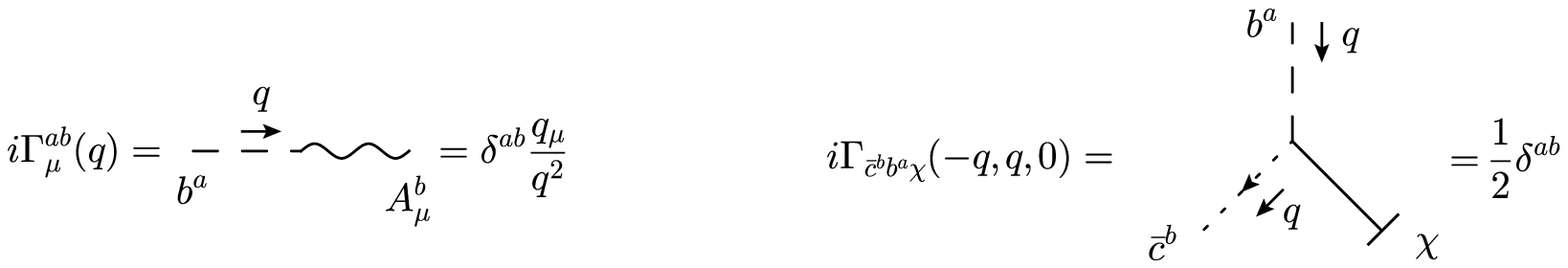}}
\caption{\label{fig:chi-Feyn-rules}Feynman rules for the $b$-sector. Notice that, due to the $b$ equation~\noeq{b-eq}, there are no possible quantum correction to these Feynman rules.}
\end{figure}

Now observe that the $b$-equation
\begin{equation}
\frac{\delta\Gamma}{\delta b^a}=\partial^\mu A^a_\mu-\xi b^a+\frac12\bar c^a\chi,
\label{b-eq}
\end{equation}
implies that the $b$ dependence is confined at tree-level, and so will the mixed $bA$ propagator~$\Delta_\mu$ (and any vertex involving the $b$ field for that matter). 
Thus, beyond tree-level the only non-trivial propagators will be
\begin{align}
i\Delta_{\mu\nu}^{ab}(q)&=-i\delta^{ab}\left[P_{\mu\nu}(q)\Delta(q^2)+\xi\frac{q_\mu q_\nu}{q^2}\right];& iD^{ab}(q)&=i\delta^{ab}\frac{F(q^2)}{q^2},
\end{align}
where $F(q^2)$ is the  ghost dressing function.
The Feynman rules for vertices involving fields and/or antifields can be found in~\cite{Binosi:2008qk}; they need to be supplemented with one more rule, describing the coupling of the $\chi$ source to a $b$ and a $\bar c$ fields, which can be read off directly from~\1eq{b-eq}:
\begin{equation}
i\Gamma_{\bar c^bb^a\chi}(-q,q,0)=\frac12\delta^{ab}.
\end{equation}
As already mentioned, this vertex will not receive quantum corrections, and will be completely fixed by its tree-level value given above. 
The Feynman rules involving the Nakanishy-Lautrup multiplier $b$ are summarized in~\fig{fig:chi-Feyn-rules}.

In addition, the Faddeev-Popov equation
\begin{equation}
\frac{\delta\Gamma}{\delta\bar c^a}+\partial^\mu\frac{\delta\Gamma}{\delta A^{*a}_\mu}=\frac12\chi b^a,
\label{FPEq}
\end{equation}
implies that beyond tree-level the dependence on $\bar c$ of the vertex functional $\Gamma$ can only be realized 
through the combination $\widetilde A^{*a}_\mu=A^{*a}_\mu+\partial_\mu c^a$; indeed, if $\Gamma=\Gamma[\widetilde A^*]$ one has
\begin{equation}
\frac{\delta\Gamma}{\delta\bar c^a(x)}=\int\!\diff^4y\,\frac{\delta\Gamma}{\delta\widetilde A^b_\mu(y)}\frac{\delta\widetilde A^b_\mu(y)}{\delta\bar c^a(x)}=\int\!\diff^4y\,\frac{\delta\Gamma}{\delta\widetilde A^b_\mu(y)}\partial_\mu^y\delta(x-y)=-\partial^x_\mu\frac{\delta\Gamma}{\delta\widetilde A^b_\mu(x)}=-\partial^x_\mu\frac{\delta\Gamma}{\delta A^b_\mu(x)},
\end{equation}
with the last step due to the linearity of the field transformation employed.
Making then the change of variable $A^*_\mu\to\widetilde{A}^*_\mu$, and introducing the reduced functional $\widetilde\Gamma$ through
\begin{equation}
\widetilde\Gamma=\Gamma-\int\!\diff^{4}x\,\left[b^a\partial^\mu A^a_\mu-\frac\xi2(b^a)^2\right],
\label{rem}
\end{equation}
one can restrict the sum over fields appearing in the rhs of~\1eq{NId} to the pairs $(A^a_\mu,\widetilde A^{*a}_\mu)$ and $(c^a,c^{*}_a)$ alone. In the NI analysis carried out in this work we have use only `tilded' quantities, and therefore suppressed this symbol everywhere. Incidentally, notice that it is~\1eq{rem} that implies the tree-level result $\Gamma^{(0)}_{A^a_\mu A^b_\nu}(q)=iq^2\delta^{ab}P_{\mu\nu}(q)$.

The final form of the NI used is then written as 
\begin{align}
\left.\frac{\partial\Gamma}{\partial\xi}\right\vert_{\chi=0}&=\left.\int\!\diff^{4}x\,\left(
\frac{\delta\Gamma}{\delta A^{*a}_\mu}\frac{\delta^2\Gamma}{\partial\chi\delta A_a^\mu}
-\frac{\delta^2\Gamma}{\partial\chi\delta A^{*a}_\mu}\frac{\delta\Gamma}{\delta A_a^\mu}
-\frac{\delta^2\Gamma}{\partial\chi\delta c^*_a}\frac{\delta\Gamma}{\delta c^a}
-\frac{\delta\Gamma}{\delta c^*_a}\frac{\delta\Gamma}{\partial\chi\delta c^a}\right)\right\vert_{\chi=0}.
\label{NId-final}
\end{align}

Using the technique developed in~\cite{Binosi:2012st}, one can write the complete solution to the NI above~\cite{Quadri:2014jha}. Rewriting~\1eq{NId-final} as\footnote{The sign differences with respect to~\cite{Quadri:2014jha} are due to the different conventions used. In particular, our Yang-Mills action is obtained from the one of~\cite{Quadri:2014jha} through the replacements: $\bar c\to-\bar c$, $b\to-b$, $c^*\to-c^*$ and $\alpha\to-\xi$ (which also implies $\theta\to-\chi$ when introducing the doublet partner of the gfp parameter).}
\begin{align}
\left.\frac{\partial\Gamma}{\partial\xi}\right\vert_{\chi=0}&=\left.\int\!\diff^{4}x\,\left(
\frac{\delta\Psi}{\delta A^{*a}_\mu}\frac{\delta\Gamma}{\delta A_a^\mu}
-\frac{\delta\Psi}{\delta A^{a}_\mu}\frac{\delta\Gamma}{\delta A_a^{*\mu}}
+\frac{\delta\Psi}{\delta c^a}\frac{\delta\Gamma}{\delta c^*_a}
-\frac{\delta\Psi}{\delta c^*_a}\frac{\delta\Gamma}{\delta c^a}
\right)\right\vert_{\chi=0};& \Psi\equiv\frac{\partial\Gamma}{\partial\chi},
\end{align}
its full solution is  given by~\cite{Quadri:2014jha}
\begin{equation}
\Gamma=\sum_{n\ge0}\frac1{n!}\xi^n\Gamma_n;\qquad\Gamma_n=\left.\left[\Delta_\Psi^n\Gamma_0\right]\right\vert_{\xi=0},
\label{gensol}
\end{equation}
where $\Gamma_0=\left.\Gamma\right\vert_{\xi=0}$ is the vertex functional in the Landau gauge, and in the $R_\xi$ gauges the Lie operator $\Delta_\Psi$ reads
\begin{equation}
\Delta_\Psi X =\int\!\diff^{4}x\,\left(
\frac{\delta X}{\delta A^{a}_\mu}\frac{\delta\Psi}{\delta A_a^{*\mu}}
+\frac{\delta X}{\delta A^{*a}_\mu}\frac{\delta\Psi}{\delta A_a^\mu}
+\frac{\delta X}{\delta c^*_a}\frac{\delta\Psi}{\delta c^a}
+\frac{\delta X}{\delta c^a}\frac{\delta\Psi}{\delta c^*_a}\right)
+\frac{\partial X}{\partial\xi}.
\end{equation}
If $\xi\ll1$ one can linearize~\1eq{gensol}; the coefficient of the linear term $\Gamma_1$ is then obtained by applying the Lie operator on the Landau vertex functional $\Gamma_0$. As the latter does not depend on the gfp $\xi$, within the linear approximation one has  
\begin{equation}
\partial_\xi\Gamma=\Gamma_1,
\label{linear}
\end{equation}
with
\begin{align}
\Gamma_1&=\left.\int\!\diff^{4}x\,\left(
\frac{\delta \Gamma_0}{\delta A^{a}_\mu}\frac{\delta\Psi}{\delta A_a^{*\mu}}
+\frac{\delta \Gamma_0}{\delta A^{*a}_\mu}\frac{\delta\Psi}{\delta A_a^\mu}
+\frac{\delta \Gamma_0}{\delta c^*_a}\frac{\delta\Psi}{\delta c^a}
+\frac{\delta \Gamma_0}{\delta c^a}\frac{\delta\Psi}{\delta c^*_a}\right)\right\vert_{\xi=0}\nonumber \\
&=\int\!\diff^{4}x\,\left(
\frac{\delta\Gamma_0}{\delta A^{*a}_\mu}\frac{\delta^2\Gamma_0}{\partial\chi\delta A_a^\mu}
-\frac{\delta^2\Gamma_0}{\partial\chi\delta A^{*a}_\mu}\frac{\delta\Gamma_0}{\delta A_a^\mu}
-\frac{\delta^2\Gamma_0}{\partial\chi\delta c^*_a}\frac{\delta\Gamma_0}{\delta c^a}
-\frac{\delta\Gamma_0}{\delta c^*_a}\frac{\delta\Gamma_0}{\partial\chi\delta c^a}\right).
\end{align}
We then see that the approximation employed in this work  
on the full NIs are equivalent to differentiating~\1eq{linear} with respect to a ghost and an antighost~[\1eq{NId-last-1}], or two gluon fields~[\1eq{NId-gl-1}].


\end{document}